\def\year{2019}\relax
\documentclass[letterpaper]{article} %DO NOT CHANGE THIS
\usepackage{aaai19}  %Required
\usepackage{times}  %Required
\usepackage{helvet}  %Required
\usepackage{courier}  %Required
\usepackage{url}  %Required
\usepackage{graphicx}  %Required
\usepackage{natbib}
\usepackage{amsmath, amssymb}

\usepackage{xcolor}

\usepackage{multirow}
\usepackage[title]{appendix} % appendix

\newcommand{\vx}{\mathbf{x}}
\newcommand{\vX}{\mathbf{X}}
\newcommand{\vY}{\mathbf{Y}}

\newcommand{\vC}{\mathbf{C}}
\newcommand{\qE}{\mathbb{E}}

\newcommand{\LL}{\mathcal{L}}
\newcommand{\XX}{\mathcal{X}}
\newcommand{\YY}{\mathcal{Y}}
\newcommand{\argmin}{\arg\,\min}

\begin{document}

\title{Play as You Like: Timbre-enhanced Multi-modal Music Style Transfer}
%\author{Anonymous AAAI Submission\\
%Paper ID 2259\\
%}
\author{Chien-Yu Lu,\textsuperscript{1}\thanks{The first two authors are with equal contribution.}
Min-Xin Xue,\textsuperscript{1*}
Chia-Che Chang,\textsuperscript{1}
Che-Rung Lee,\textsuperscript{1}
Li Su\textsuperscript{2}\\
%\author{AuthorThree},\textsuperscript{3}
%\author{AuthorFour},\textsuperscript{4}
%\author{AuthorFive}, \textsuperscript{5}\\
\textsuperscript{1}Department of Computer Science, National Tsing-Hua University, Hsinchu, Taiwan\\
\textsuperscript{2}Institute of Information Science, Academia Sinica, Taipei, Taiwan\\
\{j19550713, liedownisok, chang810249\}@gmail.com, cherung@cs.nthu.edu.tw, lisu@iis.sinica.edu.tw
}

\maketitle
\begin{abstract}
Style transfer of polyphonic music recordings is a challenging task when considering the modeling of diverse, imaginative, and reasonable music pieces in the style different from their original one. To achieve this, learning stable multi-modal representations for both domain-variant (i.e., style) and domain-invariant (i.e., content) information of music in an unsupervised manner is critical. In this paper, we propose an unsupervised music style transfer method without the need for parallel data. Besides, to characterize the multi-modal distribution of music pieces, we employ the Multi-modal Unsupervised Image-to-Image Translation (MUNIT) framework in the proposed system. This allows one to generate diverse outputs from the learned latent distributions representing contents and styles. Moreover, to better capture the granularity of sound, such as the perceptual dimensions of timbre and the nuance in instrument-specific performance, cognitively plausible features including mel-frequency cepstral coefficients (MFCC), spectral difference, and spectral envelope, are combined with the widely-used mel-spectrogram into a timber-enhanced multi-channel input representation. The Relativistic average Generative Adversarial Networks (RaGAN) is also utilized to achieve fast convergence and high stability. We conduct experiments on bilateral style transfer tasks among three different genres, namely piano solo, guitar solo, and string quartet. Results demonstrate the advantages of the proposed method in music style transfer with improved sound quality and in allowing users to manipulate the output.
\end{abstract}

\section{Introduction}
\noindent The music style transfer problem has been receiving increasing attention in the past decade \citep{dai2018music}. When discussing this problem, typically we assume that music can be decomposed into two of its attributes, namely \emph{content} and \emph{style}, the former being domain-invariant and the latter domain-variant. This problem is therefore considered as to modify the style of a music piece while preserving its content. However, the boundary that distinguishing content and style is highly dynamic; different objective functions in timbre, performance style or composition are related to different style transfer problems \citep{dai2018music}.
%timbre style transfer for sound morphing [], performance style transfer for performance control [], and composition style transfer for music rearrangement [] \citep{dai2018music}. 
Traditional style transfer methods based on feature interpolation \citep{caetano2011sound} or matrix factorization \citep{driedger2015let,su2017automatic} typically need a \emph{parallel} dataset containing musical notes in the target-domain style, and every note has a pair in the source domain. In other words, we need to specify the content attribute element-wisely, and make style transfer be performed in a supervised manner. Such restriction highly limits the scope that the system can be applied. To achieve higher-level mapping across domains, recent approaches using deep learning methods such as the generative adversarial networks (GAN) \citep{GAN} allow a system to learn the content and style attributes directly from data in an unsupervised manner with extra flexibility in mining the attributes relevant to content or style \citep{Ulyanov2016,Bohan2017,wu2018singing,DBLP:journals/corr/abs-1801-01589,haque2018conditional,mor2018universal}. 

Beyond the problem of unsupervised domain adaptation, there are still technical barriers concerning realistic music style transfer applicable for various kinds of music. % the style transfer problem of , with the demands to generate diverse, imaginative and reasonable outputs, . 
First, previous studies can still hardly achieve multi-modal and non-deterministic mapping between different domains. However, when we transfer a piano solo piece into guitar solo, we often expect the outcome of the guitar solo to be \emph{adjustable}, perhaps with various fingering styles, brightness, musical texture, or other sound quality. %A music style transfer system should transfer one input music piece into different versions of outputs in the target domain.
Second, the transferred music inevitably undergoes degradation of perceptual quality such as severely distorted musical timbre; this indicates the need of a better representation for timbre information. Although many acoustic correlates of timbre have been verified via psychoacoustic experiments \citep{grey1977multidimensional, alluri2010exploring,caclin2005acoustic} and also been used in music information retrieval \citep{lartillot2008matlab,peeters2011timbre}, they are rarely discussed in deep-learning-based music style transfer problems. This might be because of several reasons: some acoustic correlates are incompatible to the format of modern deep learning architectures; rawer data inputs such as waveforms and spectrograms are still preferred to reveal the strength of deep learning; and even, an exact theory of those acoustic correlates on human perception is still not clear in cognitive science \citep{siedenburg2016comparison, aucouturier2013seven}. For this issue, a recently proposed method in \citep{mor2018universal} adopts the WaveNet \citep{van2016wavenet}, the state-of-the-art waveform generator on raw waveform data to generate realistic outputs for various kinds of music with a deterministic style mapping, at the expense of massive computing power.
% In these models, it is assumed all the data share a common content space (domain-invariant), and each domain has its own domain-variant style space. The distinction between domain-invariant and domain-variant is not clear.

To address these issues, we consider the music style transfer problem as learning a multi-modal conditional distribution of style in the target domain given only one unpaired sample in the source domain. This is similar to the Multi-modal Unsupervised Image-to-Image Translation (MUNIT) problem, where a principled framework proposed in \citep{huang2018munit} is employed in our system. During training, cognitively plausible timbre features including mel-frequency cepstral coefficients (MFCC), spectral difference, and spectral envelope, all designed to have the same dimension with mel-spectrogram, are combined together into a multi-channel input representation in the timbre space. Since these features have close-form relationship with each other, we introduce a new loss function, named \emph{intrinsic consistency loss}, to keep the consistency among the channel-wise features in the target domain. Experiments show that with such extra conditioning on the timbre space, the system does achieve better performance in terms of content preservation and sound quality than those using only the spectrogram. Moreover, comparing to other style transfer methods, the proposed multi-modal method can stably generate diverse and realistic outputs withs improved quality. Also, in the learned representations, some dimensions that disentangle timbre can be observed. Our contributions are two-fold: 
\begin{itemize}
\item We propose an unsupervised  multi-modal music style transfer system for one-to-many generation. To the best of our knowledge, this have not been done before in music style transfer. The proposed system further allows music style transfer from scratch, without massive training data.
\item We design multi-channel timbre features with the proposed intrinsic consistency loss to improve the sound quality for better listening experience of the style-transferred music. Disentanglement of timbre characteristics in the encoded latent space is also observed. 
\end{itemize}
%Based on multi-modal representations for the domain variant and domain-invariant information, we can produce diverse music pieces. 2) We can linearly interpolate selected dimension on style code to generate a series of music pieces. 3) We can train a multi-modal style-transferred model from scratch without massive training data.

% 可能要提更多 FB 那篇 paper 的缺點?
% 1. Not from scratch traing, using WaveNet pre-trained model with massive audio data
% 2. high computing cost

% 條列式?
% Contribution:
% 	1. We can generate diverse synthetic music pieces.
% 	2. We can select specific dimension to do interpolation.
% 	3. We can directly train a style-transfer model from scratch without massive training data.

%Based on the GAN, we adopt the show that combining both sides can make improved result, with a hope to bridge the gap between music informatics and perceptual science. 
%[Our proposed system does allow the use of timbre in deep learning and doing this does achieve better performance] %This is a compromise that makes sense is to pick the relevant feature verified by psychoacoustics.
%The major components to be transferred, such as timbre and playing techniques of instruments, are better understood as a multi-dimensional object, and can be more efficiently modeled in multiple data representations, even when using deep learning. 

\section{Related Works}
\subsection{Generative Adversarial Networks}

% Introduce GANs here
Since its invention in \citep{GAN}, the GAN has shown amazing results in multimedia content generation in variant domains \citep{YuZWY17, GwakCGCS17, LiLY017}. A GAN comprises two core components, namely the generator and the discriminator. The task of the generator is to fool the discriminator, which distinguishes real samples from generated sample. 
This loss function, named \emph{adversarial loss}, is therefore implicit and is defined only by the data.
Such a property is particularly powerful for generation tasks. 

\subsection{Domain Adaptation}
% Introduce MUNIT here
Recent years has witnessed considerable success in unsupervised domain adaptation problems without parallel data, such as image colorization \citep{LarssonMS16, ZhangIE16} and image enhancement \citep{Chen:2018:DPE}. 
Two of the most popular methods that achieve unpaired domain adaptation could be the CycleGAN \citep{DBLP:journals/corr/ZhuPIE17} and the Unsupervised Image-to-Image Translation Networks (UNIT) \citep{DBLP:journals/corr/LiuBK17} framework, the former introduce the cycle consistency loss to train with unpaired data and the other is to learn a joint distribution of images in different domains.
However, most of these transfer models are based on a deterministic or one-to-one mapping. Therefore, these models are unable to generate diverse outputs when given the data from source domain. One of the earliest attempts on multi-modal unsupervised translation could be \citep{ZhuZPDEWS17}, which aims at capturing the distribution of all possible outputs, that means, a one-to-many mapping that maps a single input into multiple outputs. To handle multi-modal translation, two possible methods are: adding random noise to the generator, or adding dropout layer into the generator for capturing the distribution of outputs. However, these methods still tend to generate similar outputs since the generator is easy to ignoring random noise and additional dropout layers. In this paper, we use a disentangled representation framework, MUNIT \citep{huang2018munit}, for generating high-quality and high-diversity music pieces with unpaired training data. 
\subsection{Music Style Transfer}

The music style transfer problem has been investigated for decades. Broadly speaking, the music being transferred can be either audio signals or symbolic scores \citep{dai2018music}. In this paper, we focus on the music style transfer of audio signals, where its domain-invariant \emph{content} typically refer to the structure established by the composer (e.g., mode, pitch, or dissonance)\footnote{Although the instrumentation process is usually done by the composer, especially in Western classical music, we presume that the timbre (i.e., the instrument chosen for performance) is determined by the performer.}, and its domain-variant \emph{style} refers to the interpretation
of the performer (e.g., timbre, playing styles, expression).

With such abundant implications of content and style, the music style transfer problem encompasses extensive application scenarios, including audio mosaicking \citep{driedger2015let}, audio antiquing \citep{valimaki2008digital,su2017automatic}, and singing voice conversion \citep{kobayashi2014statistical,wu2018singing}, to name but a few. 
Recently, motivated by the success of image style transfer \citep{DBLP:conf/cvpr/GatysEB16}, 
using deep learning for music or speech style transfer on audio signals has caught wide attention. These solutions can be roughly categorized into two classes. The first class takes spectrogram as input and feeds it into convolutional neural networks (CNN), recurrent neural networks (RNN), GAN or autoencoder \citep{haque2018conditional, donahue2018synthesizing}. Cycle consistency loss has also been applied for such features \citep{wu2018singing, hosseini2018multi}. The second class takes raw waveform as input and feed it into autoregressive models such as WaveNet \citep{mor2018universal}. Unlike the classical approaches, the deep learning approaches pay less attention to the level of signal processing, and tends to overlook timbre-related features that are psychoacoustically meaningful in describing music styles. One notable exception is \citep{DBLP:journals/corr/abs-1801-01589}, which took the deviation of temporal and frequency energy envelopes respectively from the style audio into the loss function of the network, and demonstrated promising results.

\section{Data Representation} \label{Data_Representation}
\noindent We discuss the audio features before introducing the whole framework of the proposed system. We set two criteria of choosing features for our system input. First, all the features can be of the same dimension, so as to facilitate a CNN-based multi-channel architecture, where one feature occupy one input channel. In other words, the channel-wise features represent the colors of sound; this is similar to the case of image processing, where three colors (i.e., R, G, and B) are also taken as channel-wise input. Second, the chosen features should be related to music perception or music signal synthesis. The features verified to be highly correlated to one or more attributes of musical timbre through perceptual experiments are preferred more. As a result, we consider the following four data representations: 1) mel-spectrogram, 2) mel-frequency cepstral coefficients (MFCC), 3) spectral difference, and 4) spectral envelope.

Consider an input signal $\vx:=\vx[n]$ where $n$ is the index of time. Give a $N$-point window function $\mathbf{h}$ for the computation of the short-time Fourier transform (STFT): 
\begin{equation}
\vX[k,n]:=\sum^{N-1}_{m=0} \vx[m+nH]\mathbf{h}[m]e^{-\frac{j2\pi km}{N}}\,.
\end{equation}
where $k$ is the frequency index. The sampling rate is $f_s=22.05$ kHz. We consider the \emph{power spectrogram} of $\vx$ being the $\gamma$-power of the magnitude part of the STFT, namely $|\vX|^{\gamma}$. In this paper we set $\gamma=0.6$, a value that well approximate the perceptual scale based on the Stevens power law \citep{stevens1957psychophysical}. The mel-spectrogram $\bar{\vX}[f,n]:=\mathbf{M}|\vX|^{\gamma}$ is the power spectrogram mapped into the mel-frequency scale with a filterbank. The filterbank $\mathbf{M}$ has 256 overlapped triangular filters ranging from zero to 11.025 kHz, and the filters are equally-spaced in the mel scale: $\text{mel} := 2595\log_{10}(f/700+1)$.
MFCC is represented as the discrete cosine transform (DCT) of the mel-spectrum:
\begin{equation}
\vC[q,n] := \sum^{F-1}_{f=0} \tilde{\vX}[f,n]\cos\left[\frac{\pi}{N}\left(f+\frac{1}{2}\right)q\right]\,.
\end{equation}
where $q$ is the cepstral index and $F=256$ is the number of frequency bands. 
The MFCC has been one of the most widely used audio feature ranging from a wide diversity of tasks including speech recognition, speaker identification, music classification, and many others. Traditionally, only the first few coefficients of the MFCC are used, as these coefficients are found relevant to timbre-related information. High-quefrency coefficients are then related to pitch. In this work, we adopt all coefficients for end-to-end training. 

The \emph{spectral difference} is a classic feature for musical onset detection and timbre classification. It is highly relevant to the attack in the attack-decay-sustain-release (ADSR) envelope of a note. The spectral difference is represented as 
\begin{equation}
\Delta\tilde{\vX}[f,n]:=\mathrm{ReLU}(\tilde{\vX}[f,n+1]-\tilde{\vX}[f,n])
\end{equation}
where ReLU refers to a rectified linear unit that discards the energy-decreasing parts in the time-frequency plane. The accumulation of spectral difference over the frequency axis is the well-known \emph{spectral flux} for musical onset detection.

The \emph{spectral envelope} $\vY$ can be loosely estimated through the inverse DCT of the first $\eta$ elements of the MFCC, which represents the slow-varying counterpart in the spectrum:
\begin{equation}
\vY[f,n] := \sum^{\eta}_{q=0} \vC[q,n]\cos\left[\frac{\pi}{N}\left(q+\frac{1}{2}\right)f\right]\,,
\end{equation}
where $\eta$ is the \emph{cutoff cepstral index}. In this paper we set $\eta=15$. The spectral envelope has been a well-known factor in timbre and is widely used in sound synthesis [].
These data representations emphasize different aspects of timbre, and at the same time able to act as a channel for joint learning.

\section{Proposed Method}

Consider the style transfer problem from two domains $\XX$ and $\YY$. $x\in\XX$ and $y\in\YY$ are two samples from $\XX$ and $\YY$, respectively. Assume that the latent spaces of the two domains are partially shared: each $x$ is generated by a content code $c\in\mathcal{C}$ shared by both domains and a style code $s\in\mathcal{S}$ in the individual domain. Inferring the marginal distributions of $c$ and $s$, namely $p(c)$ and $p(s)$, respectively, therefore allows one to achieve \emph{one-to-many} mapping between $\XX$ and $\YY$. This idea was first proposed in the MUNIT framework \citep{huang2018munit}. To further improve its performance and to adapt to our problem formulation, we make two extensions.  
First, to stabilize the generation result and speed up the convergence rate, we adopt the Relativistic average GAN (RaGAN) \citep{DBLP:journals/corr/abs-1807-00734} instead of the for the conventional GAN component for generation. Second, considering the relation between the channel-wise timbre features, we introduce the \emph{intrinsic consistency loss} to pertain the relation between the output features. 
	\begin{figure*}[ht]
        \centering
        \includegraphics[width=\linewidth]{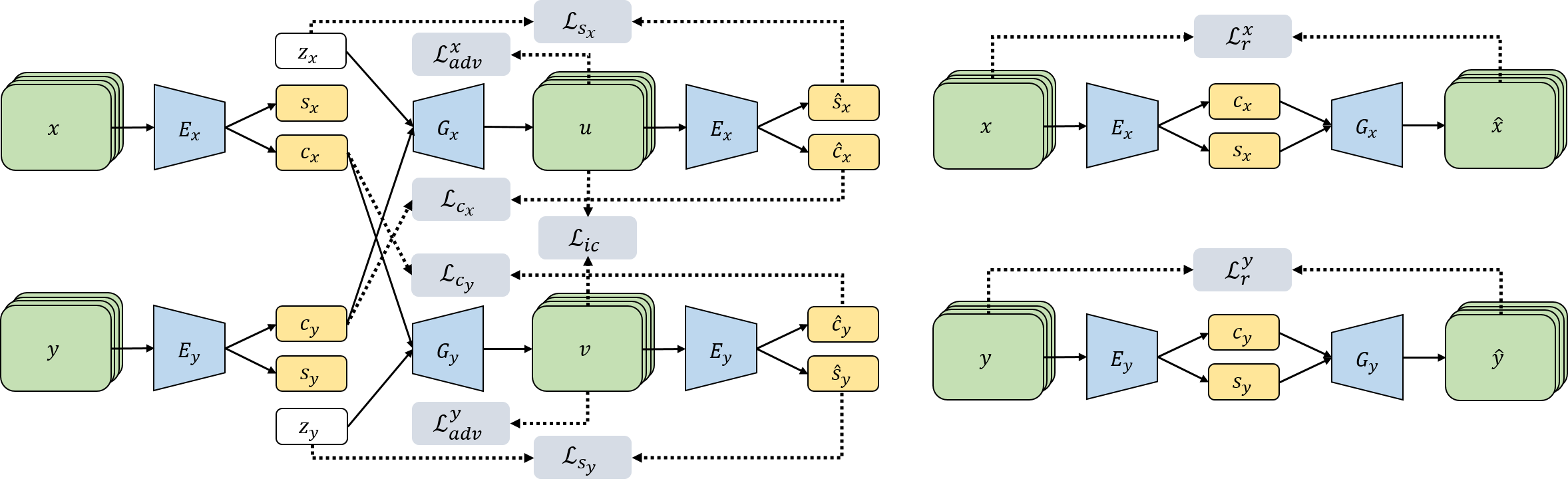}
        \caption{The proposed multi-modal music style transfer system with intrinsic consistency regularization $\LL_{ic}$. Left: cross-domain architecture. Right: self-reconstruction.}
        \label{figure:model-architecture}
	\end{figure*}
 
\subsection{Overview}
Fig. \ref{figure:model-architecture} conceptually illustrates the whole multi-mdoal music style transfer architecture. It contains encoders $E$ and generators $G$ for domains $\XX$ and $\YY$, namely $E_\XX$, $E_\YY$, $G_\XX$, and $G_\YY$.\footnote{Since the transfer task is bilateral, we will ignore the subscript if we do not specifically mention $\XX$ or $\YY$ domains. For example, $G$ refers to either $G_\XX$ or $G_\YY$} $E$ encodes a music piece into a style code $s$ and a content code $c$. $G$ decodes $c$ and $s$ into the transferred result, where $c$ and $G$ are from different domains and $s$ in the target domain is sampled from a Gaussian distribution $z\in\mathcal{N}(0,1)$. For example, the process $v=G_\YY(c_x, s_y)$ where $s_y\in\mathcal{N}(0,1)$ transfer $x$ in domain $\XX$ to $v$ in domain $\YY$. Similarly, the process transferring $y$ in domain $\YY$ to $u$ in domain $\XX$ is also shown in Fig. \ref{figure:model-architecture}.

The system has two main networks, cross-domain translation and within-domain reconstruction, as shown in the left and the right of Fig. \ref{figure:model-architecture}, respectively. The cross-domain translation network uses GANs to match the distribution of the transferred features to the distribution of the features in the target domain. It means, discriminators $D$ should distinguish the transferred samples from the ones truly in the target domain, and $G$ needs to fool $D$ by capturing the distribution of the target domain.

By adopting the Chi-Square loss \citep{DBLP:conf/iccv/MaoLXLWS17} in the GANs, 
the resulting adversarial loss, $\LL_{adv}$, is represented as:

\begin{align}
\LL_{adv} &= \LL_{adv}^x + \mathcal{L}_{adv}^y \nonumber\\
&= \qE_{c_y \sim p(c_y), z \sim \mathcal{N}}[(D_\XX(G_\XX(c_y, z)))^2]\nonumber\\
&+ \qE_{x}[(D_\XX(x)-1)^2] \nonumber\\
&+\qE_{c_x \sim p(c_x), z \sim \mathcal{N}}[(D_\YY(G_\YY(c_x, z))^2]\nonumber\\
&+ \qE_{y}[(D_\YY(y)-1)^2],
\label{equation:adversarial_loss}
\end{align}

where $p(c_y)$ is a marginal distribution from which $c_y$ is sampled. 
Besides, we expect that the content code of a given sample should remain the same after cross-domain style transfer. This is done by minimizing the content loss ($\LL_{c}$): 
\begin{equation}
\mathcal{L}_{c} = \mathcal{L}_{c_{x}} + \mathcal{L}_{c_{y}} = \vert c_y - \hat{c}_{x} \vert_1 + \vert c_x - \hat{c}_{y} \vert_1,
\label{equation:content_loss}
\end{equation}
where $|\cdot|$ is the $l_1$-norm, $c_y$ ($c_x$) is the content code before style transfer, and $\hat{c}_x$ ($\hat{c}_y$) is the content code after style transfer. Similarly, we also expect the style code of the transferred result to be the same as the one sampled before style transfer. This is done by minimizing the style loss $\LL_s$:
\begin{equation}
\LL_{s} = \LL_{s_{x}} + \LL_{s_{y}} = \vert z_x - \hat{s}_x \vert_1 + \vert z_y - \hat{s}_y \vert_1,
\label{equation:style_loss}
\end{equation}
where $\hat{s}_x$ and $\hat{s}_y$ are the transferred style codes, and $z_x$ and $z_y$ are two input style codes sampled from $\mathcal{N}(0,1)$.

Finally, the system also incorporates self-reconstruction mechanism, as shown in the right of Fig. \ref{figure:model-architecture}. For example, $G_\XX$ should be able to reconstruct $x$ from the latent codes $(c_x, s_x)$ that $E_\XX$ encodes. The reconstruction loss is

\begin{equation}
\LL_{r} = \LL_{r}^x + \LL_{r}^y = \vert x - \hat{x} \vert_1 + \vert y - \hat{y} \vert_1,
\end{equation}
where $\hat{x}$ and $\hat{y}$ are the reconstructed features of $x$ and $y$, respectively.

\subsection{RaGAN}

One of our goals is to translate music pieces into the target domain with improved sound quality. 
To do this, we adopt the recently-proposed Relativistic average GAN (RaGAN) \citep{DBLP:journals/corr/abs-1807-00734} as our GAN training methodology to generate high quality and stable outputs. RaGAN is different from other GAN architectures in that in the training stage, the generator not only captures the distribution of real data, but also decreases the probability that real data is real. 
The RaGAN discriminator is designed as
\begin{equation}
D(x)= \\
\begin{cases}
             \sigma(Q(x) - \qE_{x_f\sim\mathbb{Q}}\, Q(x_f)) \text{\ if $x$ is real}\,, \\
             \sigma(Q(x) - \qE_{x_r\sim\mathbb{P}}\, Q(x_r)) \text{\ if $x$ is fake}\,,
\end{cases}
\end{equation}
where $\sigma(\cdot)$ is the sigmoid function, $Q$ is the layer before the sigmoid output layer of the discriminator, and $x$ is the input data. $\mathbb{P}$ is the distribution of real data, $\mathbb{Q}$ is the distribution of fake data. $x_r$ and $x_f$ denote real and fake data, respectively.

\begin{figure*}[ht]
	\centering
	\includegraphics[width=0.8\linewidth, , trim = {0cm 0cm 0cm 0.3cm}, clip]	{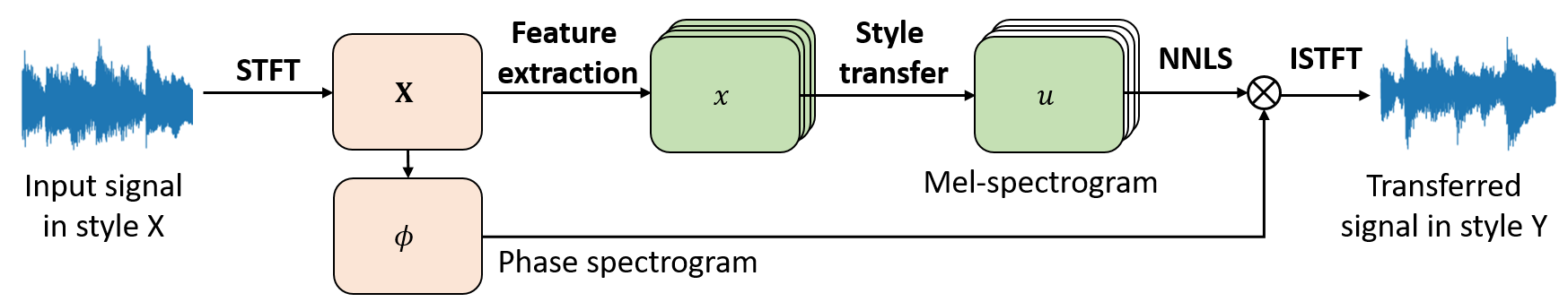}
	\caption{Illustration of pre-processing and post processing on audio signals. The power-scale spectrogram and the phase spectrogram $\Phi$ are derived from the short-time Fourier transform $\vX$. To reconstruct the generated mel-spectrogram $u_{\mathrm{ms}}$, the NNLS optimization and the original phase spectrogram $\Phi$ are used to get a stable reconstructed signal via the ISTFT.}

	\label{End-to-End-Process}
\end{figure*}

\subsection{Intrinsic Consistency Loss}

To achieve one-to-many mapping, the MUNIT framework deprecates the cycle consistency loss that is only applicable in one-to-one settings. We needs extra ways to guarantee the robustness of the transferred features. By noticing that the multi-channel features are all derived from the mel-spectrogram with closed forms, we propose a new regularization term to guide the transferred features to be with the same closed-form relation. In other words, the intrinsic relations among the channels should remain the same after style transfer. First, the MFCC channel should remain the DCT of the mel-spectrogram:

\begin{align}
\LL_{\mathrm{MFCC}} &= \LL_{\mathrm{MFCC}_{u}} + \LL_{\mathrm{MFCC}_{v}} \nonumber\\
&= \vert u_{\mathrm{MFCC}} - \mathrm{DCT}( u_{\mathrm{ms}} ) \vert_1 \nonumber\\
&+ \vert v_{\mathrm{MFCC}} - \mathrm{DCT}( v_{\mathrm{ms}} ) \vert_1 .
\end{align}
where $u_\mathrm{MFCC}$ is the transferred MFCC and $u_\mathrm{ms}$ is the transferred mel-spectrogram.
Similar loss functions can also be designed for spectral difference and spectral envelope:
\begin{align}
\LL_{\Delta} &= \LL_{\Delta}^u + \LL_{\Delta}^v \nonumber\\
&= \vert u_{\Delta} - \Delta u_{\mathrm{ms}} \vert_1 + \vert v_{\Delta} - \Delta v_{\mathrm{ms}} \vert_1 .\\
% \end{align}
% \begin{align}
\LL_{\mathrm{env}} &= \LL_{\mathrm{env}}^u + \LL_{\mathrm{env}}^v \nonumber\\
&= \vert u_{\mathrm{env}} - \mathrm{IDCT}( \mathrm{DCT}(u_{\mathrm{ms}})_{:\eta}) \vert_1 \nonumber\\
&+ \vert v_{\mathrm{env}} - \mathrm{IDCT}( \mathrm{DCT}(v_{\mathrm{ms}})_{:\eta}) \vert_1 .
\end{align}
That means, the transferred spectral difference (e.g., $u_\Delta$) should remain as the spectral difference of the transferred mel-spectrogram (e.g., $\Delta u_{\mathrm{ms}}$). The case of spectral envelope is also similar. 
The total intrinsic consistency loss is 
\begin{equation} \label{eq:Lic}
\LL_{ic} = \lambda_{\mathrm{MFCC}}\LL_{\mathrm{MFCC}}+\lambda_\Delta\LL_\Delta+\lambda_{\mathrm{env}}\LL_{\mathrm{env}}\,,
\end{equation}
and the full objective function $\LL$ of our model is
\begin{align}
\underset{E_{\XX}, E_{\YY}, G_{\XX}, G_{\YY}}{\min}\,\,\underset{D_{\XX}, D_{\YY}}{\max}\mathcal{L}(E_{x}, E_{y}, G_{x}, G_{y}, D_{x}, D_{y})\nonumber\\
=\LL_{adv} + \lambda_{c} \LL_{c} + \lambda_{s} \LL_{s} + \lambda_{r} \LL_{r} + \LL_{ic},\label{eq:full_objective_function}
\end{align}
where $\lambda_{adv}$, $\lambda_{S}$ and $\lambda_{recon}$ are hyper-parameters to reconstruction loss.

\subsection{Signal Reconstruction}
The style-transferred music signal is reconstructed from the mel-spectrogram and the phase spectrogram $\Phi$ of the input signal. This is done in the following steps. First, since the mel-spectrogram $\bar{\vX}$ is nonnegative, we can convert it back to a linear-frequency spectrogram through the mel-filterbank $\mathbf{M}$ using the nonnegative least square (NNLS) optimization:
\begin{equation}
\vX^* = \underset{\vX}{\argmin}\|\bar{\vX}-\mathbf{M}\vX\|^2_2\quad\text{subject to }\vX\succeq 0\,.
\end{equation}

The resulting magnitude spectrum is therefore $\hat{\vX}:=\vX^{*(1/\gamma)}$. Then, the complex-valued time-frequency representation $\hat{\vX}e^{j\Phi}$ is processed by the inverse short-time Fourier transform (ISTFT), and the final audio is obtained. The process dealing with waveforms is illustrated in Fig. \ref{End-to-End-Process}.

\subsection{Implementation details} 
The adopted networks are mostly based on the MUNIT implementation except for the RaGAN in adversarial training. 
The model is optimized by adam, with the batch size being one, and with the learning rate and weight decay rate being both 0.0001.
The regularization parameters in (\ref{eq:Lic}) and (\ref{eq:full_objective_function}) are: $\lambda_r=10$, $\lambda_s=\lambda_c=1$, and $\lambda_{\mathrm{MFCC}}=\lambda_\Delta=\lambda_{\mathrm{env}}=1$. The sampling rate of music signals is $f_s=22.05$ kHz. The window size and hop size for STFT are 2048 and 256 samples, respectively. The dimension of the style code is 8.

\section{Experiment and Results}

In the experiments, we consider two music style transfer tasks using the following experimental data: 
\begin{enumerate}
\item Bilateral style transfer between classical piano solo (Nocturne Complete Works performed by Vladimir Ashkenazy) and classical string quartet (Bruch's Complete String Quartet).
\item Bilateral style transfer between popular piano solo and popular guitar solo (data of both domains consists in 34 piano solos (8,200 seconds) and 56 guitar solos (7,800 seconds) covered by the pianists and guitarists on YouTube. Please see supplementary materials for details).
\end{enumerate}
In brief, there are four subtasks in total: piano to guitar (P2G), guitar to piano (G2P), piano to string quartet (P2S), and string quartet to piano (S2P). 

For each subtask, we evaluate the proposed system in two stages, the first being the comparison to baseline models and the second the comparison to baseline features. For the two baseline models, we consider CycleGAN \citep{DBLP:journals/corr/ZhuPIE17} and UNIT \citep{DBLP:journals/corr/LiuBK17}, which are both competitive unsupervised style transfer networks. Note that the two baseline models allow only one-to-one mapping. For the features, we consider using mel-spectrogram only (MS), mel-spectrogram and MFCC (MC), and all four features (ALL). For simplicity, we do not exhaust all possible combinations of these settings. Instead, we consider the following five cases: 
CycleGAN-MS, UNIT-MS, MUNIT-MS, MUNIT-MC, and MUNIT-ALL. These cases suffice the comparison on both feature and model.

Subjective tests were conducted to evaluate the style transfer system from human's perspective. For each subtask, one input music clip is transferred using the above five settings. CycleGAN and UNIT both generate one output sample, and for MUNIT-based methods, we randomly select three style codes in the target domain and obtain three output samples. This results in a huge amount of listening samples, so we split the test into six different questionnaires, three of them comparing models and the other three three comparing features. By doing so, only one out of the three MUNIT-based output needs to be selected in a questionnaire. A participant only needs to complete one randomly selected questionnaire to finish one subjective test. 

In each round, a subject first listens to the original music clip, then its three style-transferred versions using different models (i.e., CycleGAN, UNIT, MUNIT) or different features (i.e., MS, MC, ALL). For each transferred version, the subject is asked to score three problems from 1 (low) to 5 (high). The three problems are:

\begin{figure}[t]
	\centering
	\includegraphics[width=\linewidth]	{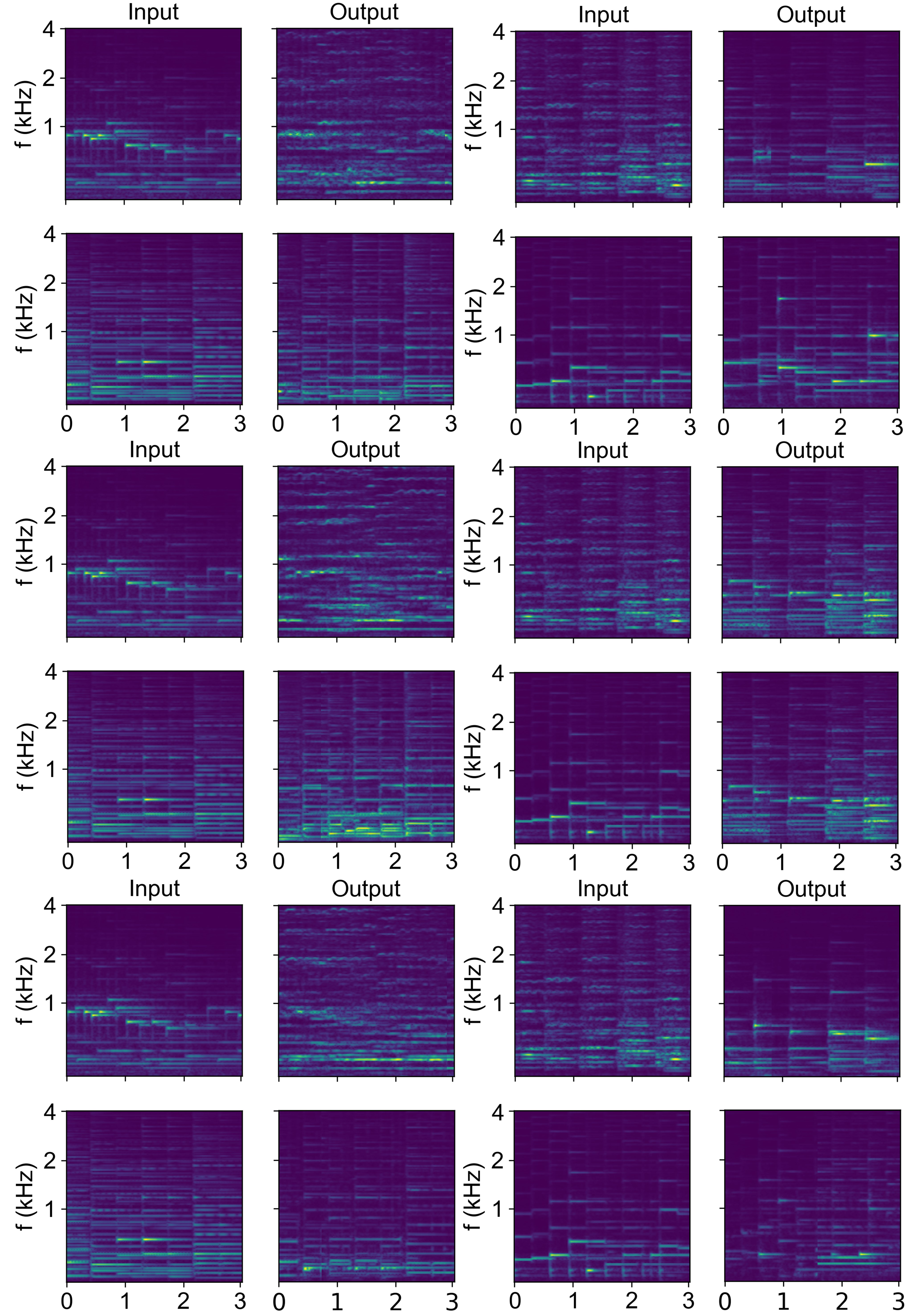}
	\caption{Comparison of the input (original) and output (transferred) mel-spectrograms for CycleGAN-MS (the upper two rows), UNIT-MS (the middle two rows), and MUNIT-MS (the lower two rows). The four subtasks demonstrated in every two rows are: P2S (upper left), S2P (upper right), P2G (lower left), and G2P (lower right).} 
	\label{VisualizedSameples:CycleGAN_UNIT_MUNIT-1v}
\end{figure}

\begin{enumerate}
\item Success in style transfer (ST): how well does the style of the transferred version match the target domain, 
\item Content preservation (CP): how well does the content of the transferred version match the original version, and 
\item Sound quality (SQ): how good is the sound.
\end{enumerate}
After the scoring process, the subject is asked to choose the best and the worst version according to her/his personal view on style transfer. This part is a preference test.

\begin{figure}[t]
	\centering
	\includegraphics[width=\linewidth]	{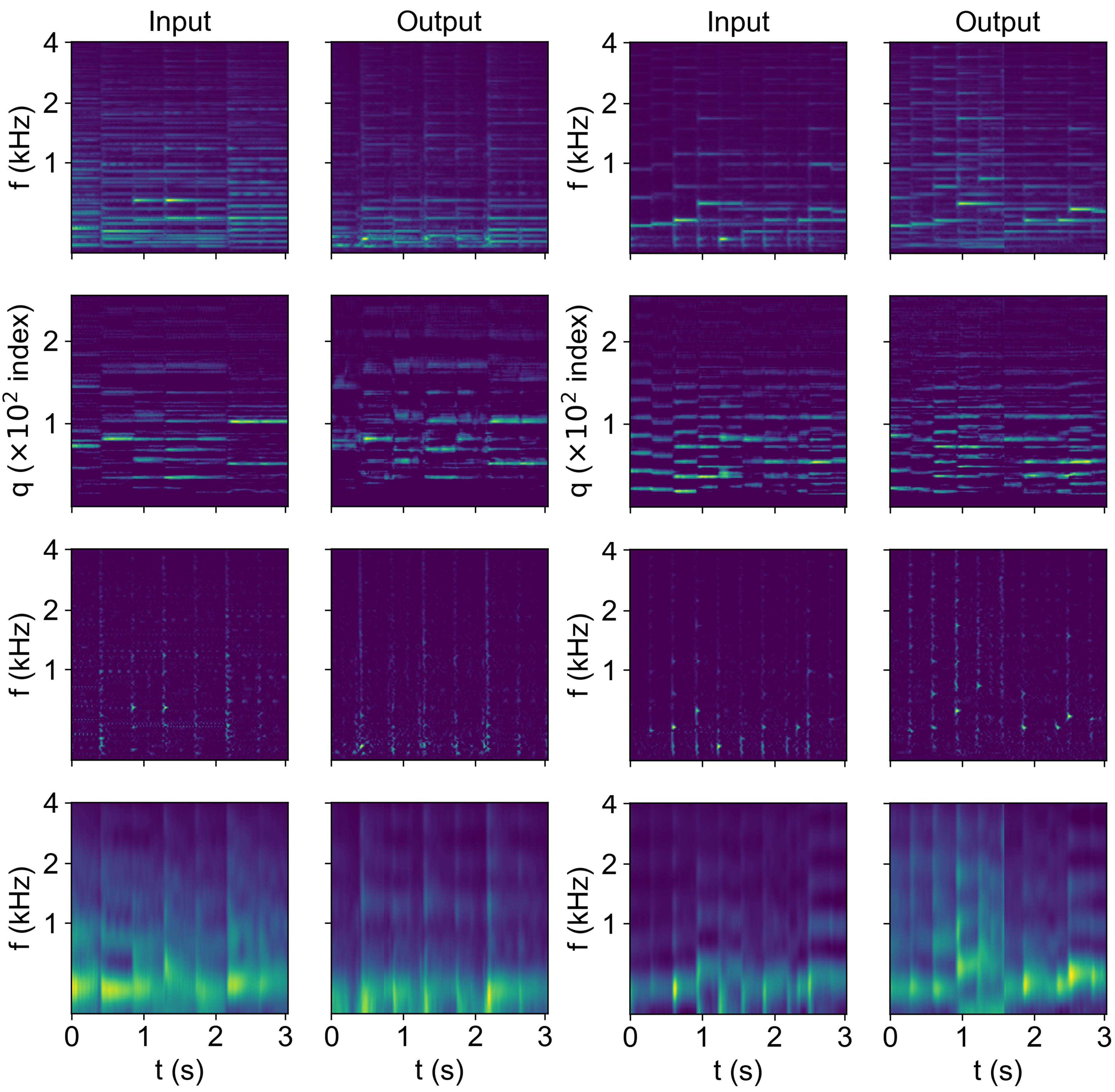}
	\caption{Illustration of the input (original) and output (transferred) feature using MUNIT-ALL of on P2G (the left two columns) and G2P (the right two columns). From top to bottom: mel-spectrogram, MFCC, spectral difference, and spectral envelope.}
	\label{VisualizedSameples:MUNIT-4-P2G}
\end{figure}

\begin{table*}[t]
\caption{The mean opinion score (MOS) of various style transfer tasks and settings. From top to bottom: CycleGAN-MS, UNIT-MS, MUNIT-MS, MUNIT-MC, MUNIT-ALL. See the supplementary material for details about the details of evaluation.} % Each transfer occupies three columns, each entry is the average score that the subject evaluate on success in style transfer, content preservation, and sound quality. }% The scores are give respectively to 4 transfers. While the average scoring show a big-trend over genres.
\label{Table:Subjective_Test_Score}
\small{
\begin{tabular}{|c|c|ccc|ccc|ccc|ccc|ccc|}
\hline
\multicolumn{2}{|c|}{Task} & \multicolumn{3}{c|}{P2G}                      & \multicolumn{3}{c|}{G2P}                      & \multicolumn{3}{c|}{P2S}                      & \multicolumn{3}{c|}{S2P}                      & \multicolumn{3}{c|}{Average}                  \\ \hline
Model        & Feature     & ST            & CP            & SQ            & ST            & CP            & SQ            & ST            & CP            & SQ            & ST            & CP            & SQ            & ST            & CP            & SQ            \\ \hline
CycleGAN     & MS          & 2.89          & \textbf{4.27} & 2.56          & 2.66          & \textbf{4.17} & 2.57          & 2.85          & 3.51          & 2.33          & 3.21          & \textbf{4.01} & 3.10          & 2.90          & \textbf{3.99} & 2.64          \\ %\hline
UNIT         & MS          & 2.85          & 4.07          & 2.80          & 2.57          & 3.83          & 2.20          & 2.83          & \textbf{3.62} & 2.28          & 3.39          & 3.90          & 2.88          & 2.91          & 3.85          & 2.54          \\ %\hline
MUNIT        & MS          & 2.97          & 3.98          & 2.64          & \textbf{3.06} & 3.91          & 2.48          & \textbf{2.88} & 3.45          & \textbf{2.43} & 3.55          & 3.56          & 2.88          & \textbf{3.12} & 3.72          & 2.61          \\ %\hline
MUNIT        & MC          & 3.30          & 4.07          & \textbf{3.14} & 2.80          & 3.56          & 2.42          & 2.77          & 3.32          & 2.27          & 3.47          & 3.44          & 2.92          & 3.09          & 3.60          & 2.69          \\ %\hline
MUNIT        & ALL         & \textbf{3.55} & 4.12          & 3.13          & 2.95          & 4.02          & \textbf{2.97} & 2.12          & 3.11          & 1.93          & \textbf{3.76} & 3.70          & \textbf{3.25} & 3.09          & 3.74          & \textbf{2.82} \\ \hline
\end{tabular}
}
\end{table*}

\begin{figure}[t]
	\centering
	\includegraphics[width=\linewidth, trim = {0.4cm 0.2cm 0.3cm 0.3cm}, clip]	{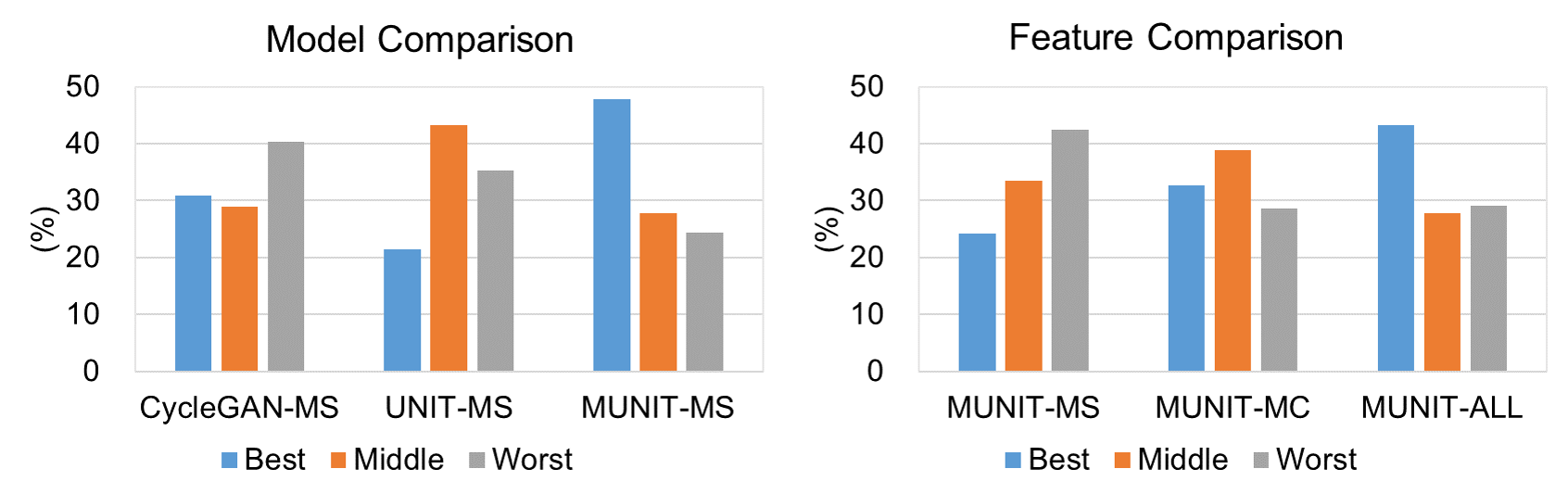}
	\caption{Results of the preference test. Left: comparison of models. Right: comparison of features. The y-axis is the ratio that each setting earns the best, middle, or the worst ranking from the listeners. 
    }
	\label{ABC_TEST_ALL}
\end{figure}

\begin{figure*}[ht]
	\centering
	\includegraphics[width=0.9\linewidth, trim = {0cm 0.5cm 0cm 1.5cm}, clip]	{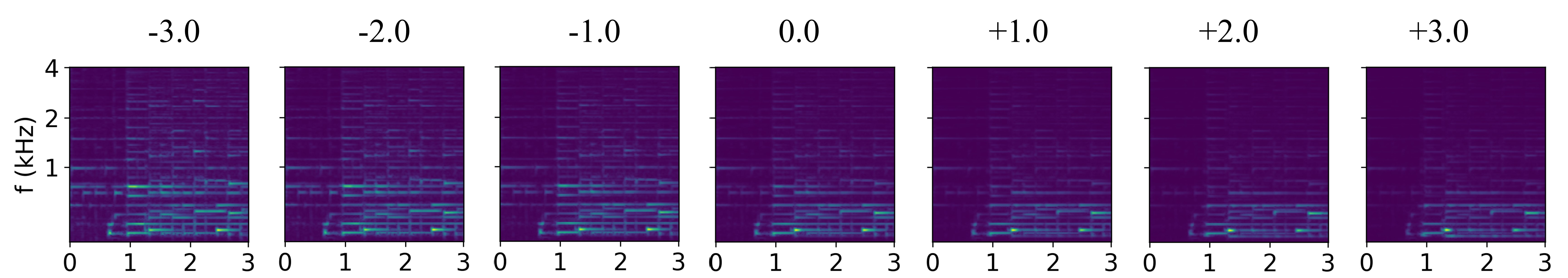}
	\caption{Converted mel-spectrograms from a piano music clip in the P2G task with the 6th dimension of the sampled style code varying from -3 to 3. The horizontal axis refers to time. Audio samples are available in the supplementary material.}
	\label{style_interpolation_dim6}
\end{figure*}

\subsection{Subjective Evaluation}
\label{Subjective_Evaluation} % audio_style_transfer <- this paper has similar subsection
Table \ref{Table:Subjective_Test_Score} shows the Mean Opinion Scores (MOS) of the listening test collected from 182 responses. %90 responses on model comparison and 92 responses on feature comparison.  
First, by comparing the three models, we can see that CycleGAN performs best in content preservation after domain transfer, possibly because of the strength of the cycle consistency loss in matching the target domain directly at the feature level.

On the other hand, MUNIT outperforms the other two models in terms of style transfer and sound quality. Second, by comparing the features, we can see that using ALL features outperforms others by 0.1 in the average sound quality score. For content preservation and style transfer, however, the number of feature is rather insensitive. 
While MUNIT-based methods get the highest scores in style transfer, which shows learning a multi-modal conditional distribution better generates realistic style-transfered output, we can't see the relation between multi-channel features and style transfer quality. However, the sound quality evaluation shows that MUNIT-ALL conducts the best sound quality.

The above results indicate an unsurprising trade-off between style transfer and content preservation. The overall evaluation of listeners' preference on those music style transfer systems could be better seen from the preference test result. The results are shown in Fig. \ref{ABC_TEST_ALL}. For the comparison of models, up to 48\% of listeners view MUNIT-MS as the best, and only 24\% of listeners views it as the worst. On the other side, CycleGAN-MS gets the most ``worst'' votes and MNUIT-MS gets the least. 
For the comparison of features, 43\% of the listeners view MUNIT-ALL as the best, and at the same time 42\% of the listeners view MUNIT-MS as the worst. These results demonstrate the superiority of the proposed method over other baselines. 
\subsection{Illustration of Examples}

Fig. \ref{VisualizedSameples:CycleGAN_UNIT_MUNIT-1v} compares the input and output mel-spectrograms among different models and tasks. From the illustrations one may observe that all the models generate some characteristics related to the target domain. For example, we observe that in the P2S task, there are vibrato notes in the output, and in the P2G task, the high-frequency components are suppressed. 
More detailed feature characteristics can be seen in Fig. \ref{VisualizedSameples:MUNIT-4-P2G} where all the four features in an P2G task are shown. For the output in guitar solo style, one may further observe longer note attacks shown in the spectral difference, and less high-frequency parts in spectral envelope, both of which are indeed characteristics of guitar. 

\subsection{Style Code Interpolation}
We then investigate how a specific dimension of the style code can affect the generation result. Fig. \ref{style_interpolation_dim6} shows a series of P2G examples with interpolated style codes. For a selected style code $z\in\mathcal{N}(0, 1)$, 
we linearly interpolate the 6th dimension of $z$, $z[6]$, with a value from -3 to 3, and generate a series of music pieces based on these modified style code. Interestingly, results show that when $z[6]$ increases, the high-frequency parts decreases. In this case, $z[6]$ can be related to some timbre features such as \emph{spectral centroid} or \emph{brightness}. 
This phenomena indicates that some of the style code elements do disentangle the characteristics of timbre.

\section{Conclusion}
We have presented a novel method to transfer a music pieces into multiple pieces in another style. We have shown that the multi-channel features in the timbre space and the regularization of the intrinsic consistency loss among them improve the sound quality of the transferred music pieces. The multi-modal framework also match the target domain distribution better than previous approaches. In comparison to other style transfer methods, our proposed method is one-to-many, stable, and without the need of paired data and pre-trained model. The learned representation of style is also adjustable. These findings suggest further studies on disentangling timbre characteristics, utilizing the findings from psychoacoustics on the perceptual dimension of music styles, and the speeding up of the music style transfer system. Codes and listening examples of this work are announced online at: https://github.com/ChienYuLu/Play-As-You-Like-Timbre-Enhanced-Multi-modal-Music-Style-Transfer

\bibliographystyle{aaai}
\bibliography{AAAI_ref}

\begin{thebibliography}{}

\bibitem[\protect\citeauthoryear{Alluri and
  Toiviainen}{2010}]{alluri2010exploring}
Alluri, V., and Toiviainen, P.
\newblock 2010.
\newblock Exploring perceptual and acoustical correlates of polyphonic timbre.
\newblock {\em Music Perception: An Interdisciplinary Journal} 27(3):223--242.

\bibitem[\protect\citeauthoryear{Aucouturier and
  Bigand}{2013}]{aucouturier2013seven}
Aucouturier, J.-J., and Bigand, E.
\newblock 2013.
\newblock Seven problems that keep mir from attracting the interest of
  cognition and neuroscience.
\newblock {\em Journal of Intelligent Information Systems} 41(3):483--497.

\bibitem[\protect\citeauthoryear{Bohan}{2017}]{Bohan2017}
Bohan, O.~B.
\newblock 2017.
\newblock Singing style transfer.
\newblock \url{http://madebyoll.in/posts/singing_ style_transfer/}.

\bibitem[\protect\citeauthoryear{Caclin \bgroup et al\mbox.\egroup
  }{2005}]{caclin2005acoustic}
Caclin, A.; McAdams, S.; Smith, B.~K.; and Winsberg, S.
\newblock 2005.
\newblock Acoustic correlates of timbre space dimensions: A confirmatory study
  using synthetic tones.
\newblock {\em The Journal of the Acoustical Society of America}
  118(1):471--482.

\bibitem[\protect\citeauthoryear{Caetano and Rodet}{2011}]{caetano2011sound}
Caetano, M.~F., and Rodet, X.
\newblock 2011.
\newblock Sound morphing by feature interpolation.
\newblock In {\em Proc. IEEE ICASSP},  22--27.

\bibitem[\protect\citeauthoryear{Chen \bgroup et al\mbox.\egroup
  }{2018}]{Chen:2018:DPE}
Chen, Y.-S.; Wang, Y.-C.; Kao, M.-H.; and Chuang, Y.-Y.
\newblock 2018.
\newblock Deep photo enhancer: Unpaired learning for image enhancement from
  photographs with gans.
\newblock In {\em CVPR},  6306--6314.

\bibitem[\protect\citeauthoryear{Dai and Xia}{2018}]{dai2018music}
Dai, S., and Xia, G.
\newblock 2018.
\newblock Music style transfer issues: A position paper.
\newblock In {\em the 6th International Workshop on Musical Metacreation
  (MUME)}.

\bibitem[\protect\citeauthoryear{Donahue, McAuley, and
  Puckette}{2018}]{donahue2018synthesizing}
Donahue, C.; McAuley, J.; and Puckette, M.
\newblock 2018.
\newblock Synthesizing audio with generative adversarial networks.
\newblock {\em arXiv preprint arXiv:1802.04208}.

\bibitem[\protect\citeauthoryear{Driedger, Pr{\"a}tzlich, and
  M{\"u}ller}{2015}]{driedger2015let}
Driedger, J.; Pr{\"a}tzlich, T.; and M{\"u}ller, M.
\newblock 2015.
\newblock Let it bee-towards nmf-inspired audio mosaicing.
\newblock In {\em ISMIR},  350--356.

\bibitem[\protect\citeauthoryear{Gatys, Ecker, and
  Bethge}{2016}]{DBLP:conf/cvpr/GatysEB16}
Gatys, L.~A.; Ecker, A.~S.; and Bethge, M.
\newblock 2016.
\newblock Image style transfer using convolutional neural networks.
\newblock In {\em {IEEE} {CVPR}},  2414--2423.

\bibitem[\protect\citeauthoryear{Goodfellow \bgroup et al\mbox.\egroup
  }{2014}]{GAN}
Goodfellow, I.~J.; Pouget{-}Abadie, J.; Mirza, M.; Xu, B.; Warde{-}Farley, D.;
  Ozair, S.; Courville, A.~C.; and Bengio, Y.
\newblock 2014.
\newblock Generative adversarial nets.
\newblock In {\em NIPS},  2672--2680.

\bibitem[\protect\citeauthoryear{Grey}{1977}]{grey1977multidimensional}
Grey, J.~M.
\newblock 1977.
\newblock Multidimensional perceptual scaling of musical timbres.
\newblock {\em the Journal of the Acoustical Society of America}
  61(5):1270--1277.

\bibitem[\protect\citeauthoryear{Gwak \bgroup et al\mbox.\egroup
  }{2017}]{GwakCGCS17}
Gwak, J.; Choy, C.~B.; Garg, A.; Chandraker, M.; and Savarese, S.
\newblock 2017.
\newblock Weakly supervised generative adversarial networks for 3d
  reconstruction.
\newblock {\em CoRR} abs/1705.10904.

\bibitem[\protect\citeauthoryear{Haque, Guo, and
  Verma}{2018}]{haque2018conditional}
Haque, A.; Guo, M.; and Verma, P.
\newblock 2018.
\newblock Conditional end-to-end audio transforms.
\newblock {\em arXiv preprint arXiv:1804.00047}.

\bibitem[\protect\citeauthoryear{Hosseini-Asl \bgroup et al\mbox.\egroup
  }{2018}]{hosseini2018multi}
Hosseini-Asl, E.; Zhou, Y.; Xiong, C.; and Socher, R.
\newblock 2018.
\newblock A multi-discriminator cyclegan for unsupervised non-parallel speech
  domain adaptation.
\newblock {\em arXiv preprint arXiv:1804.00522}.

\bibitem[\protect\citeauthoryear{Huang \bgroup et al\mbox.\egroup
  }{2018}]{huang2018munit}
Huang, X.; Liu, M.-Y.; Belongie, S.; and Kautz, J.
\newblock 2018.
\newblock Multimodal unsupervised image-to-image translation.
\newblock In {\em ECCV}.

\bibitem[\protect\citeauthoryear{Jolicoeur{-}Martineau}{2018}]{DBLP:journals/corr/abs-1807-00734}
Jolicoeur{-}Martineau, A.
\newblock 2018.
\newblock The relativistic discriminator: a key element missing from standard
  {GAN}.
\newblock {\em CoRR} abs/1807.00734.

\bibitem[\protect\citeauthoryear{Kobayashi \bgroup et al\mbox.\egroup
  }{2014}]{kobayashi2014statistical}
Kobayashi, K.; Toda, T.; Neubig, G.; Sakti, S.; and Nakamura, S.
\newblock 2014.
\newblock Statistical singing voice conversion with direct waveform
  modification based on the spectrum differential.
\newblock In {\em INTERSPEECH}.

\bibitem[\protect\citeauthoryear{Larsson, Maire, and
  Shakhnarovich}{2016}]{LarssonMS16}
Larsson, G.; Maire, M.; and Shakhnarovich, G.
\newblock 2016.
\newblock Learning representations for automatic colorization.
\newblock In {\em Proc. ECCV, Part {IV}},  577--593.

\bibitem[\protect\citeauthoryear{Lartillot, Toiviainen, and
  Eerola}{2008}]{lartillot2008matlab}
Lartillot, O.; Toiviainen, P.; and Eerola, T.
\newblock 2008.
\newblock A matlab toolbox for music information retrieval.
\newblock In {\em Data analysis, machine learning and applications}. Springer.
\newblock  261--268.

\bibitem[\protect\citeauthoryear{Li \bgroup et al\mbox.\egroup
  }{2017}]{LiLY017}
Li, Y.; Liu, S.; Yang, J.; and Yang, M.
\newblock 2017.
\newblock Generative face completion.
\newblock In {\em CVPR},  5892--5900.

\bibitem[\protect\citeauthoryear{Liu, Breuel, and
  Kautz}{2017}]{DBLP:journals/corr/LiuBK17}
Liu, M.; Breuel, T.; and Kautz, J.
\newblock 2017.
\newblock Unsupervised image-to-image translation networks.
\newblock {\em CoRR} abs/1703.00848.

\bibitem[\protect\citeauthoryear{Mao \bgroup et al\mbox.\egroup
  }{2017}]{DBLP:conf/iccv/MaoLXLWS17}
Mao, X.; Li, Q.; Xie, H.; Lau, R. Y.~K.; Wang, Z.; and Smolley, S.~P.
\newblock 2017.
\newblock Least squares generative adversarial networks.
\newblock In {\em ICCV},  2813--2821.

\bibitem[\protect\citeauthoryear{Mor \bgroup et al\mbox.\egroup
  }{2018}]{mor2018universal}
Mor, N.; Wolf, L.; Polyak, A.; and Taigman, Y.
\newblock 2018.
\newblock A universal music translation network.
\newblock {\em arXiv preprint arXiv:1805.07848}.

\bibitem[\protect\citeauthoryear{Peeters \bgroup et al\mbox.\egroup
  }{2011}]{peeters2011timbre}
Peeters, G.; Giordano, B.~L.; Susini, P.; Misdariis, N.; and McAdams, S.
\newblock 2011.
\newblock The timbre toolbox: Extracting audio descriptors from musical
  signals.
\newblock {\em The Journal of the Acoustical Society of America}
  130(5):2902--2916.

\bibitem[\protect\citeauthoryear{Siedenburg, Fujinaga, and
  McAdams}{2016}]{siedenburg2016comparison}
Siedenburg, K.; Fujinaga, I.; and McAdams, S.
\newblock 2016.
\newblock A comparison of approaches to timbre descriptors in music information
  retrieval and music psychology.
\newblock {\em Journal of New Music Research} 45(1):27--41.

\bibitem[\protect\citeauthoryear{Stevens}{1957}]{stevens1957psychophysical}
Stevens, S.~S.
\newblock 1957.
\newblock On the psychophysical law.
\newblock {\em Psychological review} 64(3):153.

\bibitem[\protect\citeauthoryear{Su \bgroup et al\mbox.\egroup
  }{2017}]{su2017automatic}
Su, S.-Y.; Chiu, C.-K.; Su, L.; and Yang, Y.-H.
\newblock 2017.
\newblock Automatic conversion of pop music into chiptunes for 8-bit pixel art.
\newblock In {\em Proc. IEEE ICASSP},  411--415.
\newblock IEEE.

\bibitem[\protect\citeauthoryear{Ulyanov and Lebedev}{2016}]{Ulyanov2016}
Ulyanov, D., and Lebedev, V.
\newblock 2016.
\newblock Singing style transfer.
\newblock \url{https://dmitryulyanov.github.io/
  audio-texture-synthesis-and-style-transfer/}.

\bibitem[\protect\citeauthoryear{V{\"a}lim{\"a}ki \bgroup et al\mbox.\egroup
  }{2008}]{valimaki2008digital}
V{\"a}lim{\"a}ki, V.; Gonz{\'a}lez, S.; Kimmelma, O.; and Parviainen, J.
\newblock 2008.
\newblock Digital audio antiquing-signal processing methods for imitating the
  sound quality of historical recordings.
\newblock {\em Journal of the Audio Engineering Society} 56(3):115--139.

\bibitem[\protect\citeauthoryear{Van Den~Oord \bgroup et al\mbox.\egroup
  }{2016}]{van2016wavenet}
Van Den~Oord, A.; Dieleman, S.; Zen, H.; Simonyan, K.; Vinyals, O.; Graves, A.;
  Kalchbrenner, N.; Senior, A.~W.; and Kavukcuoglu, K.
\newblock 2016.
\newblock Wavenet: A generative model for raw audio.
\newblock In {\em SSW},  125.

\bibitem[\protect\citeauthoryear{Verma and
  Smith}{2018}]{DBLP:journals/corr/abs-1801-01589}
Verma, P., and Smith, J.~O.
\newblock 2018.
\newblock Neural style transfer for audio spectograms.
\newblock {\em CoRR} abs/1801.01589.

\bibitem[\protect\citeauthoryear{Wu \bgroup et al\mbox.\egroup
  }{2018}]{wu2018singing}
Wu, C.-W.; Liu, J.-Y.; Yang, Y.-H.; and Jang, J.-S.~R.
\newblock 2018.
\newblock Singing style transfer using cycle-consistent boundary equilibrium
  generative adversarial networks.
\newblock {\em arXiv preprint arXiv:1807.02254}.

\bibitem[\protect\citeauthoryear{Yu \bgroup et al\mbox.\egroup
  }{2017}]{YuZWY17}
Yu, L.; Zhang, W.; Wang, J.; and Yu, Y.
\newblock 2017.
\newblock Seqgan: Sequence generative adversarial nets with policy gradient.
\newblock In {\em AAAI},  2852--2858.

\bibitem[\protect\citeauthoryear{Zhang, Isola, and Efros}{2016}]{ZhangIE16}
Zhang, R.; Isola, P.; and Efros, A.~A.
\newblock 2016.
\newblock Colorful image colorization.
\newblock In {\em Proc. {ECCV}, Part {III}}.

\bibitem[\protect\citeauthoryear{Zhu \bgroup et al\mbox.\egroup
  }{2017a}]{DBLP:journals/corr/ZhuPIE17}
Zhu, J.; Park, T.; Isola, P.; and Efros, A.~A.
\newblock 2017a.
\newblock Unpaired image-to-image translation using cycle-consistent
  adversarial networks.
\newblock {\em CoRR} abs/1703.10593.

\bibitem[\protect\citeauthoryear{Zhu \bgroup et al\mbox.\egroup
  }{2017b}]{ZhuZPDEWS17}
Zhu, J.; Zhang, R.; Pathak, D.; Darrell, T.; Efros, A.~A.; Wang, O.; and
  Shechtman, E.
\newblock 2017b.
\newblock Toward multimodal image-to-image translation.
\newblock In {\em NIPS},  465--476.

\end{thebibliography}
\clearpage
\section{Appendices}
%\subsection{Appendix A}
\subsection{Experiment Data and Listening Examples}

The data of piano solo and guitar solo  for training the style transfer models are collected from the web. For reproducibility, we put the YouTube link of the data we used in the experiments into two playlists. The links of the playlists are as follows:

\begin{itemize}
    \item The playlist of guitar solo is at: \url{https://goo.gl/zZv9SS}%https://www.youtube.com/watch?v=oS_yNsDsbaQ&list=PLFxfNcRmI5YpNrKGThakSLJmGiZ22zSNa
    \item The playlist of piano solo is at: \url{https://goo.gl/VbA2rA}%https://www.youtube.com/watch?v=Kt_JePg86b8&list=PLFxfNcRmI5YqmEeLdUn2I63cgwHkl3Ofu
\end{itemize}
%\textcolor{red}{[Li Su: is it possible to get a shorten playlist link? (and it needs to be a permanent link) This is a minor issue; it is also fine if we cannot have it]}
% Minxin: tinyurl (or any other url shortener) claims that their links are permanent.
Besides, the listening examples of the generated style-transferred audio in the four subtasks (i.e., P2G, G2P, P2S, and S2P), along with their original version, are available online at https://goo.gl/BhHzec and the GitHub repository:
https://github.com/ChienYuLu/Play-As-You-Like-Timbre-Enhanced-Multi-modal-Music-Style-Transfer
%Source and generated audio samples. -> I think we can put the links in GutHub repository?
% dropbox: \url{https://www.dropbox.com/sh/uzc1f2tjmhu7ylm/AAC9Jo--wVA3T2s05odSzEEza?dl=0}
% playlist of style-interp: https://www.youtube.com/watch?v=q_iw0A1eljY&list=PLFxfNcRmI5YoHwAOTlSeTqGKxshCuPKZf
%-> https://goo.gl/2dxR2U
%\begin{itemize}
%    \item The playlist of P2G is at: \url{https://tinyurl.com/y8lavkyw}
%	%https://www.youtube.com/watch?v=B_5DlsJ6Rq8&list=PLFxfNcRmI5Yp1imMPJFJMazf6n51nHz36
%    \item The playlist of G2P is at: \url{https://tinyurl.com/y8at9jgw}
%	%https://www.youtube.com/watch?v=8luOgajhLY0&list=PLFxfNcRmI5Yr0nScrDQSQRWrlkXXsSJH3
%	\item The playlist of P2S is at: \url{https://tinyurl.com/yblof3a7}
%	%https://www.youtube.com/watch?v=9XHkFyuYgpo&list=PLFxfNcRmI5Yp--L3txADH5bT_x6eY4s5t
%	\item The playlist of S2P is at: \url{https://tinyurl.com/y8s2pgok}
%	%https://www.youtube.com/watch?v=4kRBlAUk-Tc&t=0s&list=PLFxfNcRmI5YqiRacwyI39xSG8LkE7w9cI
%\end{itemize}
%

%For someone who may be interested in the piano solos and guitar solos used in bilateral style transfer between piano and guitar. Below is the list of them by YouTube links:

\subsection{Further Details on Subjective Evaluation}
In the following we report further details on the subjective evaluation. Our subjective evaluation process is completed through online questionnaires. 182 people joined our subject test. 23 of them are under 20 years old, 127 of them are between 20 and 29, 21 of them are between 30 and 39, and the rest 11 ones are above 40. We did not collect the participants' gender information, but their background of music training: 58 of the participants reported themselves as professional musicians. We take the responses from these 58 subjects as the responses from musicians, and other responses as from non-musicians. 

As mentioned in the paper, we conducted two sets of experiments, one considering the comparison on models and the other on features. The former compares Cycle-MS, UNIT-MS, and MUNIT-MS, while the latter compares MUNIT-MS, MUNIT-MC, and MUNIT-ALL. That means, the setting MUNIT-MS is evaluated in both experiments. What we reported in the paper is the average result of MUNIT-MS. Though merging the two MUNIT-MS results or not do not affect our conclusion of this paper, we can still see more details when reporting them separately. It is valuable for further discussion.

Based on the above reasons, in the supplementary material we further report 1) the mean opinion scores (MOS) given separately from musicians and non-musicians and 2) the two separated MUNIT-MS results in different scenarios of comparison, as listed in Table \ref{Table:Subjective_Test_with_Background}. Table \ref{Table:Subjective_Test_with_Background} indicates that, first, musicians tend to rate lower scores than non-musicians do in answering the questions in the subjective tests. Second, for most of the questions, the best settings the musicians and non-musicians selected are consistent. For example, in the P2G subtask, we may see from the P2G columns that both musicians and non-musicians evaluate the MUNIT model to outperform others in ST and SQ, and the CycleGAN is the best in CP. Similar observation can also be found in G2P and P2S subtasks. 

Second, the two MUNIT-MS results are different. More specifically, the MOS in feature comparison is lower than in the other, since MUNIT-MS is `relatively' inferior to the other two features, and relatively superior to the other two models. This implies the users' bias when comparing one setting under different scenario. 

Finally, there are a subtle disagreement between between musicians and non-musicians when comparing different features: on average, musicians tend to say MC is better than ALL in ST. This is mainly affected by the fact that musicians is much more sensitive than non-musicians to the low quality of the P2S results.

\begin{table*}[th]
\caption{The mean opinion score (MOS) of various style transfer tasks, models, features (Feat), and with consideration of subjects' background (BG). The ``Y/N'' on the third column represents whether the subjects report themselves as professional musicians. The upper part of the Table lists the responses of model comparisons, where we have 31 musicians and 59 non-musicians. On the other hand, the lower part collects the responses of feature comparisons, where we have 27 musicians and 65 non-musicians. Therefore, we have two sets of resulting scores for the setting MUNIT-MS.
The highest scores from two are in bold font, as we can see, the best settings the musicians and non-musicians selected are consistent for most of the questions.}
\label{Table:Subjective_Test_with_Background}
\small{
\begin{tabular}{|@{\hspace{0.08cm}}c@{\hspace{0.08cm}}|@{\hspace{0.08cm}}c@{\hspace{0.08cm}}|@{\hspace{-0.08cm}}c@{\hspace{-0.08cm}}|ccc|ccc|ccc|ccc|ccc|}
\hline
\multicolumn{3}{|c|}{Task}                                                & \multicolumn{3}{c|}{P2G}                                                    & \multicolumn{3}{c|}{G2P}                                                    & \multicolumn{3}{c|}{P2S}                                                    & \multicolumn{3}{c|}{S2P}                                                    & \multicolumn{3}{c|}{Average}                                                \\ \hline
Model                     & Feat             & \multicolumn{1}{c|}{BG} & \multicolumn{1}{c|}{ST} & \multicolumn{1}{c|}{CP} & \multicolumn{1}{c|}{SQ} & \multicolumn{1}{c|}{ST} & \multicolumn{1}{c|}{CP} & \multicolumn{1}{c|}{SQ} & \multicolumn{1}{c|}{ST} & \multicolumn{1}{c|}{CP} & \multicolumn{1}{c|}{SQ} & \multicolumn{1}{c|}{ST} & \multicolumn{1}{c|}{CP} & \multicolumn{1}{c|}{SQ} & \multicolumn{1}{c|}{ST} & \multicolumn{1}{c|}{CP} & \multicolumn{1}{c|}{SQ} \\ \hline
\multirow{2}{*}{CycleGAN} & \multirow{2}{*}{MS} & Y                       & 2.68                    & \textbf{4.06}           & 2.52                    & 2.58                    & \textbf{3.84}           & 2.68                    & 2.94                    & 3.29                    & 2.10                    & 3.19                    & \textbf{4.00}           & \textbf{3.19}           & 2.85                    & \textbf{3.80}           & 2.62                    \\
                          &                     & N                       & 3.02                    & \textbf{4.39}           & 2.61                    & 2.71                    & \textbf{4.32}           & 2.53                    & 2.85                    & 3.63                    & 2.47                    & 3.20                    & \textbf{3.98}           & \textbf{3.03}           & 2.94                    & \textbf{4.08}           & 2.66                    \\ \cline{1-3}
\multirow{2}{*}{UNIT}     & \multirow{2}{*}{MS} & Y                       & 2.65                    & 3.77                    & 2.68                    & 2.42                    & 3.42                    & 2.23                    & 3.00                    & 3.39                    & 2.39                    & 3.55                    & 3.97                    & 3.03                    & 2.90                    & 3.64                    & 2.58                    \\
                          &                     & N                       & 2.95                    & 4.22                    & 2.86                    & 2.64                    & 4.02                    & 2.19                    & 2.76                    & 3.75                    & 2.24                    & 3.32                    & 3.86                    & 2.81                    & 2.92                    & 3.96                    & 2.53                    \\ \cline{1-3}
\multirow{2}{*}{MUNIT}    & \multirow{2}{*}{MS} & Y                       & \textbf{3.03}           & 3.81                    & \textbf{2.77}           & \textbf{3.26}           & 3.81                    & \textbf{2.74}           & \textbf{3.13}           & \textbf{3.45}           & \textbf{2.55}           & \textbf{3.74}           & 3.74                    & 3.00                    & \textbf{3.29}           & 3.70                    & \textbf{2.77}           \\
                          &                     & N                       & \textbf{3.14}           & 4.22                    & \textbf{2.86}           & \textbf{3.34}           & 4.24                    & \textbf{2.68}           & \textbf{3.17}           & \textbf{3.83}           & \textbf{2.64}           & \textbf{3.69}           & 3.88                    & 2.98                    & \textbf{3.33}           & 4.04                    & \textbf{2.79}           \\ \cline{1-3} \hline
\end{tabular}
% tabular 2
\begin{tabular}{|c|@{\hspace{0.1cm}}c@{\hspace{0.1cm}}|@{\hspace{-0.05cm}}c@{\hspace{-0.05cm}}|ccc|ccc|ccc|ccc|ccc|}
\hline
\multicolumn{3}{|c|}{Task}                & \multicolumn{3}{c|}{P2G}                                                    & \multicolumn{3}{c|}{G2P}                                                    & \multicolumn{3}{c|}{P2S}                                                    & \multicolumn{3}{c|}{S2P}                                                    & \multicolumn{3}{c|}{Average}                                                \\ \hline
Model                  & Feat              & \multicolumn{1}{c|}{BG} & \multicolumn{1}{c|}{ST} & \multicolumn{1}{c|}{CP} & \multicolumn{1}{c|}{SQ} & \multicolumn{1}{c|}{ST} & \multicolumn{1}{c|}{CP} & \multicolumn{1}{c|}{SQ} & \multicolumn{1}{c|}{ST} & \multicolumn{1}{c|}{CP} & \multicolumn{1}{c|}{SQ} & \multicolumn{1}{c|}{ST} & \multicolumn{1}{c|}{CP} & \multicolumn{1}{c|}{SQ} & \multicolumn{1}{c|}{ST} & \multicolumn{1}{c|}{CP} & \multicolumn{1}{c|}{SQ} \\ \hline
\multirow{2}{*}{MUNIT} & \multirow{2}{*}{MS}  & Y                       & 2.48                    & 3.89                    & 2.22                    & \textbf{2.70}           & 3.81                    & 1.93                    & 2.37                    & 3.26                    & 1.96                    & 3.56                    & 3.37                    & 2.70                    & 2.78                    & 3.58                    & 2.20                    \\
                       &                      & N                       & 3.00                    & 3.88                    & 2.55                    & 2.86                    & 3.69                    & 2.40                    & 2.72                    & 3.18                    & 2.37                    & 3.34                    & 3.26                    & 2.80                    & 2.98                    & 3.50                    & 2.53                    \\ \cline{1-3}
\multirow{2}{*}{MUNIT} & \multirow{2}{*}{MC}  & Y                       & 3.07                    & 3.96                    & \textbf{3.15}           & \textbf{2.70}           & 3.63                    & 2.19                    & \textbf{2.48}           & \textbf{3.48}           & \textbf{2.07}           & 3.44                    & 3.52                    & 2.74                    & \textbf{2.93}           & 3.65                    & 2.54                    \\
                       &                      & N                       & 3.42                    & \textbf{4.12}           & 3.14                    & 2.86                    & 3.54                    & 2.52                    & \textbf{2.88}           & \textbf{3.28}           & \textbf{2.38}           & 3.49                    & 3.43                    & 3.02                    & 3.16                    & 3.59                    & 2.77                    \\ \cline{1-3}
\multirow{2}{*}{MUNIT} & \multirow{2}{*}{ALL} & Y                       & \textbf{3.37}           & \textbf{4.19}           & 2.93                    & 2.41                    & \textbf{4.04}           & \textbf{2.59}           & 1.59                    & 3.15                    & 1.48                    & \textbf{3.78}           & \textbf{3.81}           & \textbf{3.19}           & 2.79                    & \textbf{3.80}           & \textbf{2.55}           \\
                       &                      & N                       & \textbf{3.65}           & 4.11                    & \textbf{3.22}           & \textbf{3.15}           & \textbf{4.03}           & \textbf{3.12}           & 2.34                    & 3.12                    & 2.12                    & \textbf{3.75}           & \textbf{3.68}           & \textbf{3.29}           & \textbf{3.22}           & \textbf{3.73}           & \textbf{2.94}           \\ \cline{1-3} \hline
\end{tabular}
}
\end{table*}

\end{document}